\documentstyle[aaspp4,flushrt,pstricks]{article}
\tighten

\def\muller {M\"{u}ller}
\def\am {$^\prime$}
\begin{document}

\title{ABELL 754: A NON-HEAD ON COLLISION OF SUBCLUSTERS}

\author{Mark J.\ Henriksen}
\affil{Department of Physics, University of North Dakota , Grand Forks, ND
58202-7129\\
E-mail: mahenrik@plains.NoDak.edu}

\and

\author{Maxim L.\ Markevitch\altaffilmark{1}}
\affil{Department of Astronomy, University of Virginia , Charlottesville, VA 
22903-0818\\
E-mail: mlm5y@virginia.edu}

\altaffiltext{1}{Also IKI, Moscow, Russia}

\centerline{Submitted to {\em The ApJ Letters}, 1995 April 15}

\begin{abstract}
We have analyzed spatially resolved spectra of the A754 cluster of galaxies
obtained with ASCA. Through earlier observations with HEAO-1, Einstein, and
ROSAT as well as optical studies, A754 has been established as the prototype
system for a merger in progress. The combination of spectral and spatial
resolution over a broad energy band provided by ASCA has set unprecedented
constraints on the hydrodynamical effects of a cluster merger. We find
significant gas temperature variations over the cluster face, indicating
shock heating of the atmosphere during the merger. The hottest
region, $>$ 12 keV (90\% confidence), is located in the region of the
Northwest Galaxy clump though the entire region along the cluster axis
appears to be hotter than the mean cluster temperature ($\sim$9 keV). The
cool, $\le$5 keV, gas originally found with the HEAO1-A2 experiment, resides
in the exterior of the cluster atmosphere and in plume of gas we identify
with a stripped cool atmosphere of the infalling subcluster. We have also
attempted to reconstruct an iron abundance map of this merging system.
Though poorly constrained, no significant deviations of abundance from the
mean value are apparent in the individual regions.

A754 is the only cluster so far which shows the significant temperature
pattern expected in a subcluster merger, in both the ROSAT (Henry \& Briel
1995) and ASCA data, providing the first possibility to compare it with 
theoretical predictions. The cluster does not feature a hot peak accompanied
by two hot lobes perpendicular to the cluster axis, predicted by
hydrodynamic simulations of a head-on merger. The observed temperature and
surface brightness maps suggest that the two colliding subunits have missed
each other by about 1~Mpc, and are now moving perpendicular to the cluster
axis in the image plane (as, e.g., in the simulations by Evrard et al.\
1996).

\end{abstract}


\section{Introduction}

A754, a rich hot cluster of galaxies at $z=0.0541$ (Bird 1994) has become
the prototype of a merging cluster. It has been observed by every modern
X-ray observatory, and a series of papers in the last few years, based on
these data have established that there is evidence of a merger in progress.
Fabricant (1986) used the X-ray imaging data obtained by the Einstein IPC
and interpreted the elongated shape of the surface brightness distribution
as merging subclusters. Henriksen (1986) fit a $\beta$-model to the Einstein
imaging data to obtain the gas density profile and utilized a polytropic
relationship to predict gas temperature profiles which were then constrained
by the HEAO1-A2 spectra.  These data required non-isothermality in the gas
and were consistent with a monotonically decreasing temperature profile with
the hottest material in the center and the coolest on the outside. In 1993,
Henriksen analyzed HEAO1-A2 spectra along with Einstein IPC and SSS spectra
and found further evidence of non-isothermality in the gas and that the data
were also consistent with a simpler two-component model including a very hot
and a cooler gas. The lack of combined spatial and spectral resolution in
the data made the location of the thermal components ambiguous. Henry and
Briel (1995, hereafter HB) analyzed ROSAT PSPC data in the energy band
0.5--2~keV, which provided a high quality image of the cluster as well as
spectral constraints. These authors found that the hot gas is in the general
vicinity of the North-West (NW) galaxy clump identified by Zabludoff and
Zaritsky (1995), while the cluster brightness peak, in the vicinity of the
South-East (SE) galaxy group, has a lower temperature as does the outer
region of the cluster atmosphere.

Hydrodynamical simulations of the effect of the subcluster merger on the
intracluster medium predict the existence of relatively long-lived spatial
temperature variations in the post-merger cluster gas (e.g., Schindler \&
\muller\ 1993; Roettiger, Burns \& Loken 1993; Evrard et al.\ 1996). ASCA
data have spectral and spatial resolution combined with a broad energy band
which is sufficient to test these predictions and determine the evolutionary
stage of clusters, and put constraints on the physics of cluster mergers. In
this Letter, we present an analysis of spatially resolved spectra obtained
by ASCA of A754, which have provided temperature and abundance maps of the
cluster.

\section{Observations and Analysis}

A754 was observed by ASCA (Tanaka et al.\ 1994) for 15--18 ks with the SIS
(most of it in the 4-CCD mode which covers a square 22\am $\times$ 22\am\
field) and for 21 ks with the GIS. We used all four ASCA detectors in the
analysis. For reconstruction of the cluster gas temperature map, the scheme
from Markevitch (1996, and references therein) has been used. It consists of
simultaneous fitting the temperature in all chosen image regions, taking
into account the ASCA mirror scattering. The mirror Point Spread Function
was modeled using Cyg~X-1 data for energies above 2~keV (Takahashi et al.\
1995). For the energies 1.5--2~keV where these data are inaccurate, the PSF
data from the 2--3 keV interval were used after the appropriate correction
for the energy dependence of the GIS spatial resolution. Including this
energy interval in the analysis did not not change the best-fit values
noticeably while improving statistics. We chose to use only energies above
1.5~keV for the spatially-resolved analysis to avoid further extrapolation
of the PSF model, while using all energies above 0.5~keV in the overall
spectrum analysis. The PSF uncertainty of 15\% (1$\sigma$) was included in
the confidence interval calculation. There are no bright sources in the
vicinity of A754 so a possible stray light contribution (Ishisaki 1996) is
unimportant for our analysis. The background was modeled using the blank
field observations normalized according to their exposures, and an
uncertainty of such normalization of 20\% was also included. A ROSAT PSPC
image of the cluster was used as a brightness template as in Markevitch
(1996). The data from GIS and SIS were binned in 8 and 7 wide spectral bins,
respectively. The poorly calibrated 2--2.5~keV interval was excluded;
including it didn't change the best-fit values significantly but made
$\chi^2$ worse. Iron abundance and the absorbing column were assumed
constant at their Ginga and Galactic values, respectively, and varying them
didn't change the best-fit temperatures significantly.

The PSF effects are most important for the regions of low relative surface
brightness, while in many cases one can get a meaningful temperature
estimate for the brightness peak regions by fitting its spectrum directly
without accounting for the added scattered flux. We have performed such fits
including lower energies for several regions, to cross-check the results of
the general method.

\section{Results: Temperature and Abundance Maps} 

The GIS-measured luminosity of A754 in the 0.5--10~keV band is
$2.4\times10^{44}\,h^{-2}$ ergs~s$^{-1}$, where $H_0=100h$
km~s$^{-1}$~Mpc$^{-1}$. The emission-weighted temperature using all four
detectors is $8.5\pm0.5$ keV (9~keV using the energy interval of 1.5--11 keV
used for the temperature map), in agreement with the 90\%
confidence range for GINGA,
$8.6\pm0.5$ keV and HEAO1-A2, 7.8--9.0 keV. 
The abundance, 0.26$\pm$0.05, is derived fitting a single component
Raymond and Smith model with Xspec utilizing the Solar ratios
in Allen (1973). 
It is consistent with the Ginga value of 0.27$\pm$0.04, confirming
that A754 has a metallicity typical of a rich cluster.  The HEAO1-A2 values
are taken from Henriksen (1993) and the Ginga results from Arnaud (1994).

A two-dimensional temperature map is shown in Fig.~1, in which the ASCA
temperatures (grayscale) are overlaid on the ROSAT PSPC surface brightness
contours. The regions in which the temperatures were reconstructed are nine
5 arcmin boxes in the inner part, and four segments of the 16 arcmin-radius
circle in the outer part. As anticipated for a cluster undergoing merger,
the temperature distribution is significantly non-uniform: regions 1, 4, 7
and 8 along the cluster axis, are hotter than the cluster average at a
greater than 90\% confidence, while regions 10 and 13 on the outside are
significantly cooler. The obtained temperatures generally agree with 
HB's ROSAT PSPC values within the much larger ROSAT errors.  Exceptions are
our regions 6 and 12 whose ASCA temperatures are consistent with the
average, while HB's values for approximately the same regions are higher,
however at a low statistical significance. In the region approximately
corresponding to our region 2+3 which includes the cluster brightness peak,
HB detected a significantly lower temperature than ours.  We have looked
into the disagreement and found that when a smaller region centered on the
PSPC brightness peak is considered and the lower energies are included in
the analysis (without the PSF correction, see section above), our best-fit
temperature is lower, see Table~1. This indicates that the spectrum in this
region may be multicomponent, and the origin of the discrepancy is in the
higher sensitivity of PSPC to the lower-temperature component while ASCA is
picking the higher-temperature component. We have also made an estimate
without the PSF correction for some other selected regions. The region
centered on the SE galaxy group (Fabricant et al. 1986; Zabludoff \&
Zaritsky 1995), 3\am\ to the South from the PSPC peak, also appears cooler.
For a circle centered on the NW galaxy group (about 4\am\ westward from the
center of the region 8 in the temperature map), the simple analysis yields a
high temperature, 11.5~keV, in agreement with that from the full analysis.
It may be worth noticing that using the concentric annuli around the cluster
center, both with and without correcting for the PSF, would conceal the
complex temperature structure of this cluster (see Table~1) within the
central 20\am. The azimuthally averaged temperature appears to drop with
radius beyond $r\sim 10'\simeq 0.4\,h^{-1}$~Mpc, as seen from Fig.~1.

For the first time we have attempted to reconstruct a two-dimensional
abundance map of this cluster. Abundance variations may be an important
diagnostic in assessing gas mixture expected from the turbulence generated
by the subcluster merger.  To derive the abundances in the regions for which
the temperatures have been obtained while avoiding fitting an unmanageable
number of free parameters, we have fixed the temperatures at their best-fit
values obtained with constant abundances. It is a justifiable thing to do
because the ASCA energy resolution and energy coverage are sufficient to
separate fitting of the temperature and the abundance. The resulting
abundances in the individual regions are given in Table~2. The abundance
constraints are poor, due to the insufficient statistics of the ASCA
observation, with most of the abundances consistent with the average within
the 68\% confidence intervals. We will return to the abundances in the
section below.

\section{Discussion}

Summarizing the findings of ROSAT and ASCA, A754 is significantly
non-isothermal. There is a ridge of hot gas located approximately alongside
the cluster elongation axis, with the highest temperatures in the area of
the NW galaxy concentration. The X-ray brightness peak, elongated in the
direction perpendicular to the cluster axis, contains gas which is
significantly cooler than the cluster average, as are the cluster outskirts
toward North and East. The question which we would like to address using our
derived temperature distribution is the evolutionary history of A754.
Detailed hydrodynamic simulations of cluster mergers (e.g., Schindler \&
\muller\ 1993; Roettiger, Burns \& Loken 1993; Pearce et al.\ 1994) predict
that as a result of a recent or ongoing merger, the cluster should have a
strongly peaked temperature distribution with the highest temperature at the
site of the subcluster collision, accompanied by hot lobes perpendicular
to the collision axis. Briel \& Henry (1994) reported on the detection of
such hot lobes in another cluster, A2256, using ROSAT PSPC.  However, their
existence was not confirmed by ASCA (Markevitch 1996), who also failed to
detect the central temperature peak in that cluster. Using ASCA data, Arnaud
et al.\ (1994) and Markevitch et al.\ (1994) reported irregular temperature
structure in Perseus and in A2163, respectively, attributing it to the
merger effects, although the former have ignored the PSF scattering while
the statistical significance of the latter result was marginal. A754 remains
the only cluster for which there is a significant and unambiguous
indication of the irregular temperature structure expected in a merger,
providing the first possibility to compare it with theoretical predictions.

If one assumes, in line with all previous studies of A754, that the merger
in this cluster proceeds in the direction of the cluster elongation, one
expects to find a temperature structure quite different from what is
observed.  We offer two possible explanations for this. One is that unlike
in the simulations which considered highly supersonic ($\sim$ 3000
km~s$^{-1}$) mergers, the subcluster collision velocity is low and the
post-shock gas temperature, proportional to the upstream shock velocity
squared, is not significantly high. Analysis of optical data indicates that
mergers may likely occur at a lower velocity, for example, $\sim 500$
km~s$^{-1}$ for A98 (Beers, Geller \& Huchra 1982) and $\sim$ 1500
km~s$^{-1}$ for A3395 (Henriksen \& Jones 1996). The line-of-sight velocity
difference of the galaxy clumps in A754 is very low, about 100 km~s$^{-1}$
(Zabludoff \& Zaritsky 1995) though application of a simple dynamical model
by these authors results in a post-merger, relative velocity of $\sim2000$
km~s$^{-1}$. However, the observed heating along the cluster axis remains
to be explained in this scenario.

A more plausible explanation is a merger with a non-zero impact parameter.
Evrard et al.\ (1996) show a simulation of a merger which is reprinted here
as our Fig.~2 and is strikingly similar to A754.  In this scenario, the
merger proceeds in the image plane with the subunits infalling from North
and South with the impact parameter $\sim 0.5\,h^{-1}$ Mpc, as shown by the
gas velocity arrows in the left panel of Fig.~2 (gas is expected to drag
behind the dark matter during a merger.)  The hottest region is at the site
of the NW group penetrating the larger subunit, and the brightness
elongation to the South is the tail of this group.  The plume of cool gas at
the X-ray brightness peak is the stripped atmosphere, and perhaps even a
cooling flow, belonged to the subunit associated with the SE galaxy group,
its elongation (more clearly seen in the full resolution ROSAT image
presented in HB) pointing in the direction of that group's infall.

The fact that the cold gas still exists in the cluster may indicate that the
merger has not proceeded very far and this is the first encounter of the
subclusters. It is therefore interesting to see if any large-scale elemental
abundance differences exist, indicating that the gas belonged to different
subclusters and still retains its identity (if the subunit's abundances have
been different). We have averaged the abundances over the North-East
(regions 2, 3, 6, 10, 13 in Fig.~1) and the South-West (regions 4, 7, 8, 11,
12) areas of the cluster associated with the two infalling subunits
(excluding the regions 1, 5, 9), and found the abundances of $0.45
\pm 0.12$ and $0.29 \pm 0.13$ (68\% intervals), respectively. They are
consistent within the statistical errors, although there is a hint of
difference and it would be interesting to measure them with a better
accuracy.

To conclude, using ASCA data, we have found a prominent structure in the
temperature map of A754 and put some constraints on the abundances in the
different cluster regions. All of the evidence, including the X-ray image
and the optical galaxy distribution, is consistent with a non-head-on merger
proceeding with high velocity in the image plane. With such favorable
viewing angle and the merger stage, A754 is an ideal laboratory to test
various evolutionary scenarios.

\acknowledgments

MJH gratefully acknowledges support from NASA grant NAG5-2529 and the
hospitality of the ASCA GOF at GSFC where part of the analysis was
completed. MM was supported by NASA grants NAG5-2526 and NAG5-1891.

\pagestyle{empty}

\clearpage

\renewcommand{\arraystretch}{1.3}
\renewcommand{\tabcolsep}{3.5mm}
\begin{center}
\begin{tabular}{lcrr}
\hline \hline
Region & $r$, arcmin  & $kT_e$, keV & Abundance \\
\hline
PSPC peak &       1.3  & $6.1^{+1.1}_{-0.8}$ & $0.35^{+0.24}_{-0.24}$\nl
SE Galaxy Clump & 1.8  & $6.6^{+1.2}_{-1.0}$ & $0.35^{+0.35}_{-0.35}$\nl 
NW Galaxy Clump & 2.3  &$11.5^{+3.5}_{-2.2}$ & $0.35^{+0.39}_{-0.35}$\nl
Annulus 1 & 0--3.5     & $8.6^{+0.5}_{-0.6}$ & $0.28^{+0.07}_{-0.07}$\nl
Annulus 2 & 3.5--7.5   & $8.3^{+0.3}_{-0.4}$ & $0.25^{+0.02}_{-0.05}$\nl
Annulus 3 & 7.5--10.8  & $8.7^{+0.6}_{-0.5}$ & $0.27^{+0.07}_{-0.07}$\nl
\hline
\end{tabular}
\end{center}

Table 1.---Results from modeling regions without taking account of PSF,
including all energies. Errors are 68\%. The values for the brightness peaks
should be reasonably close to the actual values, while values for the annuli
may be affected by the PSF scattering and are given here for illustration
only. The annuli are centered approximately on the region 5 of Fig.~1.

\vspace{2cm}

\renewcommand{\arraystretch}{1.3}
\renewcommand{\tabcolsep}{3.5mm}
\begin{center}
\begin{tabular}{cccc}
\hline \hline
Region & Abundance & 68\% Interval \\
\hline
1  & 0.60& 0.1--0.9  \nl
2  & 0.31& 0.26--0.46 \nl
3  & 0.63& 0.1--0.8 \nl
4  & 0.00& 0--0.3  \nl
5  & 0.31& 0.18--0.41 \nl
6  & 0.12& 0--0.6 \nl
7  & 0.10& 0--1.2 \nl
8  & 0.28& 0--0.3 \nl
9  & 0.00& 0--0.3 \nl
10 & 0.89& 0.5--1.3 \nl
11 & 0.39& 0.2--0.9 \nl
12 & 0.32& 0.1--0.5  \nl
13 & 0.36& 0.1--0.6 \nl
\hline
\end{tabular}
\end{center}

Table 2.---Iron abundances in the regions of Fig.~1. Temperatures were kept
fixed at their best-fit values obtained for the uniform best-fit abundance
(see text).

\clearpage

\pspicture(0,2.5)(15,20)

\rput{0}(8,14.5){\epsfxsize=14cm
\epsffile{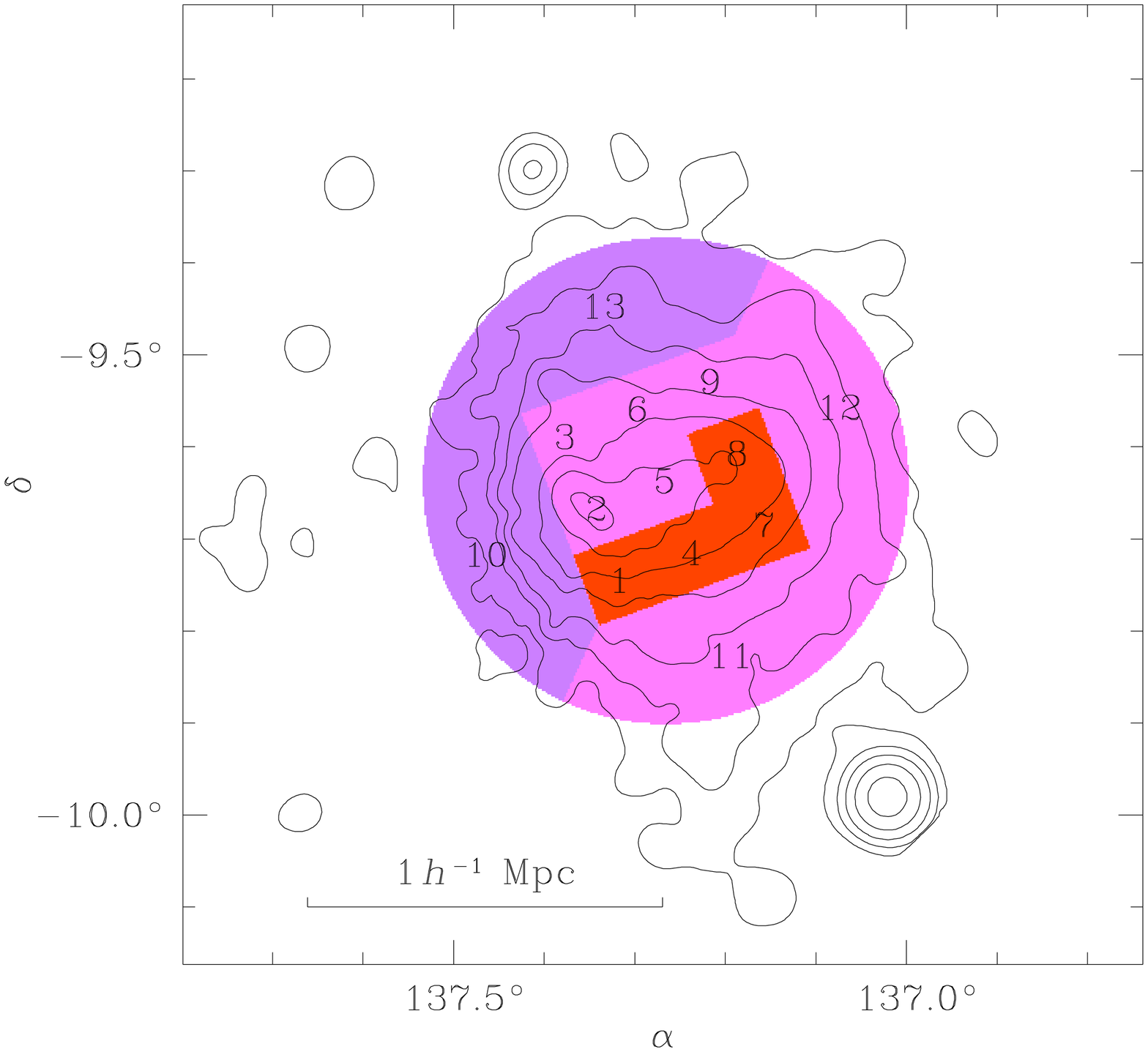}}

\rput{0}(8.39,5.9){\epsfxsize=14.7cm
\epsffile{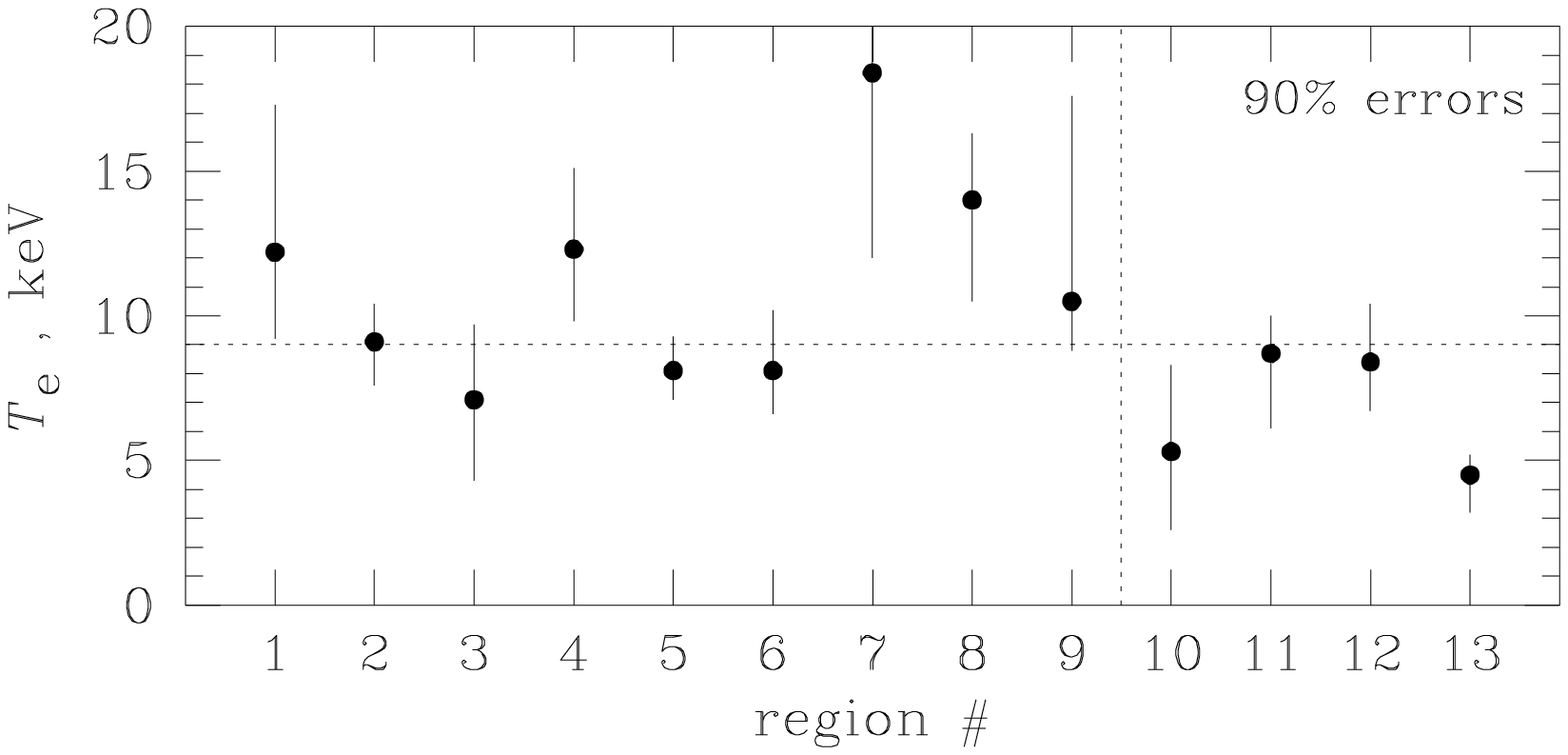}}

\rput[tl]{0}(-1,21){\small color}
\endpspicture

Fig.~1.---Temperature map of A754. Contours show the ROSAT brightness and
color shows ASCA temperatures of the cluster regions marked by their numbers
in the lower panel. The regions are nine 5 arcmin boxes in the inner part
and segments of the 16 arcmin-radius annulus in the outer part. The lower
panel shows temperature values with their one-parameter 90\% confidence
intervals for each region. Dashed line is drawn at the average cluster
temperature (9.0~keV) obtained using the same energy band, 1.5--11keV.

\clearpage

\pspicture(0,2.5)(15,20)

\rput{0}(8,14.5){\epsfxsize=14cm
\epsffile{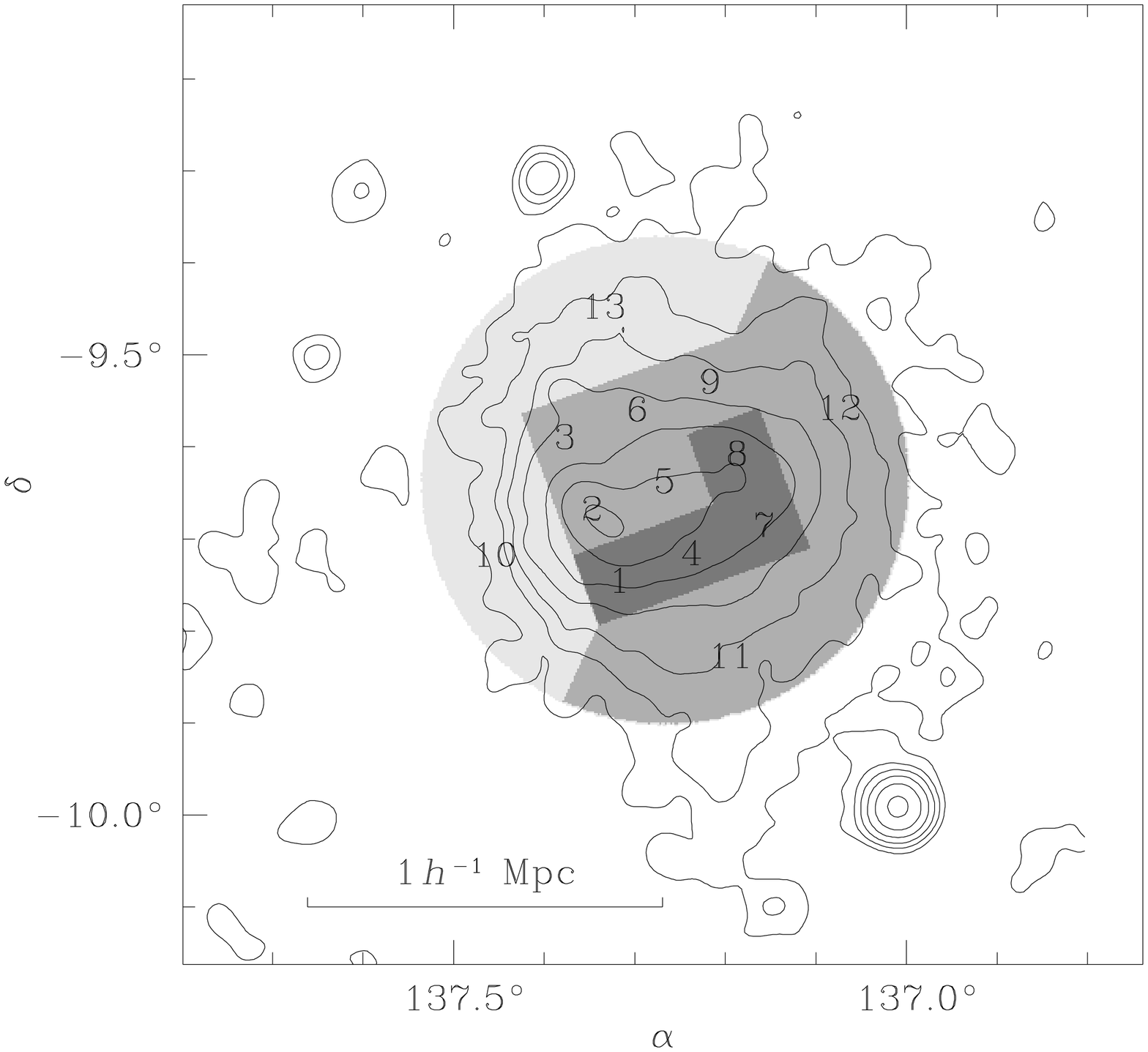}}

\rput{0}(8.39,5.9){\epsfxsize=14.7cm
\epsffile{temps.ps}}

\rput[tl]{0}(-1,21){\small grayscale}
\endpspicture

Fig.~1.---Temperature map of A754. Contours show the ROSAT brightness and
grayscale shows ASCA temperatures (darker is hotter) of the cluster regions
marked by their numbers in the lower panel. The regions are nine 5 arcmin
boxes in the inner part and segments of the 16 arcmin-radius annulus in the
outer part. The lower panel shows temperature values with their
one-parameter 90\% confidence intervals for each region. Dashed line is
drawn at the average cluster temperature (9.0~keV) obtained using the same
energy band, 1.5--11keV.

\clearpage

\pspicture(0,13)(17,20)

\rput{0}(2.5,19.2){gas velocities}
\rput{0}(7.9,19.2){X-ray brightness}
\rput{0}(13.3,19.2){gas temperature}

\rput{0}(8,16){\epsfxsize=6cm
\rotateleft{
\epsffile{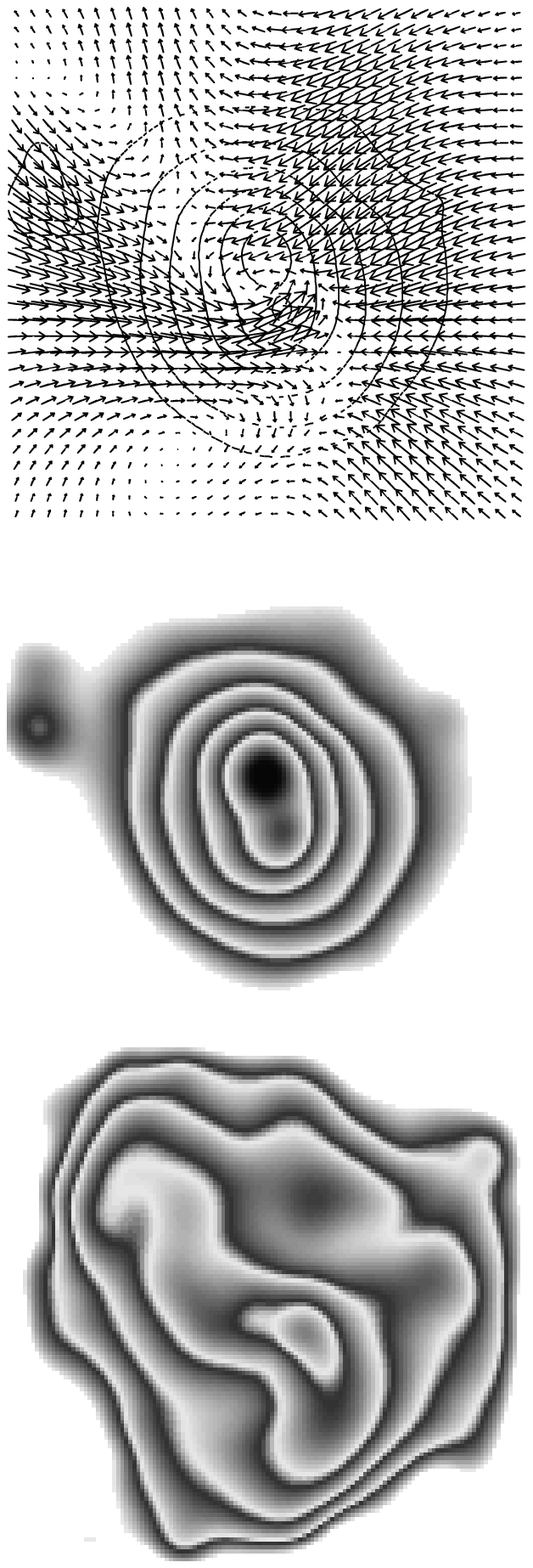}}}
\endpspicture

\vspace{1cm}
Fig.~2.---A simulated cluster from Evrard et al.\ (1996), rotated to match
the orientation of A754 (included with permission from A. Evrard.) The
brightness and temperature maps closely resemble those of A754 (Fig.~1).
Arrows in the left panel denote cluster gas velocities, overlaid on the
X-ray brightness contours. The gas generally drags behind the dark matter
during a merger. The grayscale bands in the right panel which shows the
X-ray emissivity-weighted temperature, are spaced logarithmically by
$10^{0.2}$.  The merger is proceeding in the North-South direction with a
non-zero impact parameter, which we propose as a scenario for A754.

\end{document}